\documentclass{ieeeaccess}

\usepackage{cite}
\usepackage{amsmath,amssymb,amsfonts}
\usepackage{graphicx}
\usepackage{textcomp}
\usepackage{url}
\usepackage{multirow}
\usepackage{amsthm}
\usepackage{mathrsfs}
\usepackage[table]{xcolor}
\usepackage{manyfoot}
\usepackage{booktabs}
\usepackage{braket}
\usepackage{algorithm}
\usepackage{algpseudocode}
\usepackage{listings}
\usepackage{tabularx}
\usepackage{rotating}

\usepackage{newunicodechar}
\newunicodechar{−}{-}
\newunicodechar{–}{-}
\newunicodechar{—}{-}
\newunicodechar{’}{'}
\newunicodechar{‘}{'}
\newunicodechar{“}{``}
\newunicodechar{”}{''}

\usepackage[hidelinks]{hyperref}

\def\BibTeX{{\rm B\kern-.05em{\sc i\kern-.025em b}\kern-.08em
		T\kern-.1667em\lower.7ex\hbox{E}\kern-.125emX}}

\def\BibTeX{{\rm B\kern-.05em{\sc i\kern-.025em b}\kern-.08em
    T\kern-.1667em\lower.7ex\hbox{E}\kern-.125emX}}
\begin{document}

\title{Resilience of Entanglement-Induced Coordination in Adversarial Environments: The Team-Based Quantum Sabotage Game}

\author{\uppercase{Sinan Bugu}\authorrefmark{1,2}}
\address[1]{Department of Physics and Astronomy, North Carolina State University, Raleigh,
North Carolina, USA}
\address[2]{BuQuLab Research Laboratory, Winston-Salem, North Carolina, USA}

\markboth
{Bugu: Resilience of Entanglement-Induced Coordination in Adversarial Environments: The Team-Based Quantum Sabotage Game}
{Bugu: Resilience of Entanglement-Induced Coordination in Adversarial Environments: The Team-Based Quantum Sabotage Game}

\corresp{Corresponding author: Sinan Bugu (e-mail: sbugu@ncsu.edu; sinanbugu@gmail.com ).}

\begin{abstract}
Quantum game theory extends classical strategic decision-making by incorporating quantum superposition, entanglement, and measurement-induced randomness into competitive interactions. This paper introduces a team-based Quantum Sabotage Game (QSG), in which classical and quantum-enhanced teams engage in adversarial decision-making under identical information constraints. Unlike baseline classical teams, whose members act independently, quantum teams employ entanglement-assisted coordination, generating structured correlations among decentralized actions without classical communication. We develop a formal quantum game-theoretic framework and analyze multi-agent strategies using Bell and W entangled states, benchmarked against size-equivalent classical teams. Using numerical simulations, we compare outcome distributions, correlation structure, and robustness under ideal conditions, standard quantum noise models, and a device-inspired, reproducible hardware-like noise model via the Qiskit Aer \textit{FakeKyiv} backend. While the symmetric payoff structure precludes any asymptotic increase in expected utility, multipartite entanglement, particularly W-state correlations, reshapes the finite-run joint-action distribution, producing nonclassical coordination patterns rather than an expectation-value advantage. These patterns persist under realistic noise, demonstrating that the resulting correlation signatures remain observable and differ from those produced by independent classical sampling. These results clarify the operational role of entanglement in adversarial environments, distinguishing correlation-based coordination from expectation-value advantage, and establish the Quantum Sabotage Game as a testbed for studying noise-resilient multi-agent quantum decision-making.
\end{abstract}


\doi{}
\titlepgskip=-15pt

\maketitle
\pagestyle{plain}
\thispagestyle{plain}

\section{Introduction} 

\PARstart{G}{ame} theory has long served as a foundational approach for modeling strategic decision-making in competitive and cooperative environments. Initially developed to address economic and military conflicts, it has since found applications in a broad spectrum of fields, including cybersecurity, artificial intelligence, evolutionary biology, and financial markets \cite{von1947theory, nash1950equilibrium, osborne1994course}. In cybersecurity, game-theoretic models help design adaptive defense systems that respond to evolving attack surfaces \cite{patil2018applications}, while in artificial intelligence, strategic reasoning models guide cooperative decision-making among autonomous agents. Furthermore, in biology, evolutionary game theory explains the emergence of collective optimization and adaptive behaviors in competitive ecosystems \cite{leimar2023game}, and in military contexts, game-theoretic analysis supports resource allocation, deception, and defense planning \cite{ho2022game}. Classical game theory models strategic interactions under the assumption that rational players choose deterministic or probabilistic strategies to optimize their payoffs. The rapid advancement of quantum computing has paved the way for extending these classical models into the quantum domain, giving rise to the field of quantum game theory \cite{eisert1999quantum, meyer1999quantum, piotrowski2003invitation}.

Quantum game theory introduces fundamental quantum mechanical principles, superposition, entanglement, and measurement-induced state collapse, into strategic decision making. Unlike classical players, who select a single strategy at each turn, quantum players can exist in a coherent superposition of multiple strategies, expanding the operational decision space \cite{eisert1999quantum, Flitney2002}. Moreover, quantum entanglement allows correlated strategic choices between players without direct communication, introducing a new coordination mechanism in adversarial and cooperative interactions \cite{Brunner2013, Du2020, Iqbal2009}. These quantum effects can lead to outcome statistics and correlation structures that deviate from classical mixed strategies, with differences expressed through nonclassical correlations rather than guaranteed changes in expected payoff.

Recent advances in experimental quantum computing have provided empirical validation of quantum strategic effects. Studies have demonstrated that entanglement can resolve classical dilemmas by fostering cooperation, as observed in the quantum version of the Prisoner's Dilemma \cite{eisert1999quantum, du2002entanglement, li2022prisoner,altintas2022prisoners}. Other quantum adaptations of classical games, such as the Battle of the Sexes \cite{Flitney2002, lopez2012battle} and the Quantum Colonel Blotto Game \cite{maioli2019quantization, jayanti4712364multiplayer,xu2022experimental,makram2024time, sanz2025mapping, shen2025quantum, tiago2025classical}, highlight how quantum resources reshape competitive dynamics. Experimental implementations using photonic circuits and superconducting qubits further confirm that quantum games yield measurable differences from classical counterparts \cite{schmid2010experimental}. Collectively, these results suggest that quantum game theory has implications for settings where correlation structure and decentralization matter, including adversarial AI and cybersecurity \cite{Zhao2021, Gross2020}.

Despite these advances, the study of team-based quantum games remains an open challenge. Many prior works focus on single-player or two-player quantum games, while multi-agent strategic interactions in adversarial settings introduce scaling, noise, and coordination constraints. In classical game theory, team-based sabotage games model scenarios in which competing teams allocate resources toward building, defending, and attacking opponents \cite{harsanyi1973games, levine2017deep}. Such models are widely used in cybersecurity, where defensive strategies must counteract adversarial attacks \cite{Zhao2021, kolokoltsov2022dynamic}. However, these classical sabotage games typically represent coordination using explicit communication or shared classical randomness, and they do not capture entanglement-enabled joint statistics.

This paper introduces the Team-Based Quantum Sabotage Game (QSG), an extension of classical sabotage games into the quantum domain. Two teams, a classical team and a quantum-enhanced team, engage in sabotage-only interactions under identical information constraints. Classical team members choose sabotage paths independently, without communication or coordination. The quantum team, by contrast, leverages an entangled resource state and measurement-induced randomness to generate structured correlations among decentralized actions without classical communication. In our implementation, local operations are fixed and actions are generated by a single joint measurement of the shared state, followed by a deterministic bit-to-action mapping. This produces nonclassical coordination patterns even though the symmetric payoff structure precludes an unconditional increase in expected utility.

Real-world competitive systems often involve multiple interdependent agents operating under noisy conditions. Extending quantum game theory into this multi-agent domain therefore requires both scalable entanglement structures and realistic error modeling. The QSG proposed here addresses this by integrating multi-qubit coordination, probabilistic sabotage actions, and noise-resilient correlation signatures within a unified model.

A key contribution of this work is a systematic analysis of entanglement-based coordination signatures in adversarial team-based decision-making. Prior studies have explored quantum strategies in two-player games such as the Prisoner's Dilemma and the Battle of the Sexes \cite{eisert1999quantum, Flitney2002}, but the role of multipartite entanglement in sabotage-only scoring with hidden exogenous defense has been less explored. Bugu et al. \cite{Bugu2020} showed that quantum resources can generate nonclassical coordination in adversarial settings through pseudo-telepathy games. Building on this perspective, we compare different entanglement structures (Bell and W) through their induced joint-action distributions under sabotage-only scoring with an exogenously assigned, hidden defense state. Our comparisons are structured to ensure fair evaluation: two-player classical teams against two-qubit Bell-state teams, and three-player classical teams against three-qubit W-state teams. Across these matched comparisons, W-state entanglement produces a distinct multipartite correlation structure that reshapes the finite-run outcome distribution relative to independent classical sampling. W-states have also been studied in quantum networking and multi-party quantum communication contexts \cite{Ozdemir2011, Bugu2013A, Ozaydin2014A}, motivating their use as a coordination resource in noisy multi-agent settings.

Although other multipartite states such as GHZ and Dicke states provide highly entangled configurations, they can be fragile to qubit loss and decoherence \cite{nielsen2010quantum}. In contrast, W-states retain partial entanglement even when one qubit is lost, making them natural candidates for modeling distributed decision-making under noise. This robustness has been discussed in studies of multipartite entanglement under mixing and decoherence \cite{zhu2024robustness}, further motivating our focus on W-states.

We also examine the impact of quantum noise and decoherence on coordination signatures. While many theoretical studies assume idealized quantum conditions, real hardware introduces errors that can disrupt entanglement \cite{preskill2018quantum, nielsen2010quantum}. To address this, we incorporate standard quantum noise models, including depolarizing, amplitude-damping, and bit-flip channels, and we also evaluate a hardware-calibrated noise model derived from IBM Quantum device calibration data via a reproducible Aer backend. The goal is not to claim a guaranteed payoff improvement, but to test whether correlation signatures in joint-action statistics persist under realistic noise.

In what follows, Section~\ref{sec:theoretical_model} defines the Quantum Sabotage Game and the mapping from quantum measurement outcomes to team actions. Section~\ref{sec:simulation_implementation} describes the simulation methodology, including quantum circuit construction and noise modeling. Section~\ref{sec:results} presents numerical results, focusing on outcome distributions, accumulated-score trajectories over finite horizons, and noise resilience. Section~\ref{sec:discussion} discusses implications and limitations.

Rather than asserting a universal performance advantage, this work isolates how entanglement reshapes the joint statistics of decentralized actions under identical information constraints. In particular, we show that multipartite entanglement generates structured coordination patterns that remain observable under realistic noise, even when the asymptotic expected payoff remains bounded by game symmetry.

\section{Theoretical model}\label{sec:theoretical_model}

The Quantum Sabotage Game (QSG) extends classical adversarial game formulations into the quantum domain, using entanglement and measurement-induced randomness to generate joint-action correlations among decentralized agents. The central object of study in this paper is how the shared quantum resource reshapes the joint distribution of team actions, and how that distribution changes under noise.

\subsection{Sabotage Games in Classical and Quantum Domains}

Sabotage games represent adversarial models in which players allocate actions toward undermining opponents \cite{Gross2020, Zhao2021}. These models have applications in cybersecurity, where attackers disrupt networks while defenders distribute limited security resources to minimize damage \cite{tambe2011security}. Similar formulations appear in economic competition, where agents strategically attempt to weaken rivals. In many classical models, players select discrete actions according to fixed probability rules, and coordination typically requires explicit communication or shared classical randomness.

The quantum extension introduces new correlation mechanisms. Quantum teams can employ multipartite entangled states so that their decentralized actions are sampled from a non-product joint distribution, even without classical communication. Operationally, the circuit prepares a shared state and a measurement samples a bitstring according to the Born rule. The measurement outcome then induces a correlated set of team actions through a fixed mapping. In this model, uncertainty is not used as an online adversarial weapon, but rather as an intrinsic sampling mechanism that can produce correlation patterns that are not available to independent classical strategies.

\subsection{Multi-Agent Quantum Strategies and Nash Equilibria}

A quantum Nash equilibrium (QNE), recalled here for contextual completeness, is defined as a profile of local strategies such that no agent can unilaterally improve its expected utility by modifying only its own operation while holding all other strategies fixed.
In the Eisert--Wilkens--Lewenstein (EWL) framework, each player controls a local unitary operation acting on its qubit, and the joint outcome statistics are determined by the shared entangled state and the subsequent measurement process.

Formally, let $\boldsymbol{\lambda} = \{(\theta_i,\phi_i)\}_{i=1}^{N}$ denote the set of local strategy parameters for an $N$-agent team. The expected team utility can be written as
\begin{equation}
E[U] = \sum_{\mathbf{b}\in\{0,1\}^N} P(\mathbf{b}\mid\boldsymbol{\lambda})\,U(\mathbf{b}),
\end{equation}
where $\mathbf{b}$ is the measured bitstring, $P(\mathbf{b}\mid\boldsymbol{\lambda})$ is the Born-rule probability induced by the circuit (including noise, when present), and $U(\mathbf{b})$ is the utility assigned by the game.

In this work, we do not derive closed-form quantum Nash equilibria, nor do we perform strategy-optimization sweeps over local SU(2) parameters. The equilibrium terminology is included only for contextual completeness within the EWL framework. In our simulations, strategies are fixed by state preparation and a fixed measurement-to-action mapping, and our analysis focuses on how multipartite entanglement reshapes the joint-action distribution and how this correlation structure responds to decoherence and hardware-calibrated noise.

\subsection{Quantum vs. Classical Strategy Representation}

\subsubsection{Classical Strategy Representation}
In classical sabotage games, a player's strategy $\sigma_C$ is represented as a probability distribution over two sabotage actions:
\begin{equation}
\sigma_C = (p_{S_A}, p_{S_B}), \quad \text{where} \quad p_{S_A} + p_{S_B} = 1.
\end{equation}
Each player's decision follows predefined probabilities, leading to independent or weakly correlated actions. Without communication, a classical team typically samples a product distribution across agents.

\subsubsection{Quantum Strategy Representation}

In contrast to classical sabotage actions, a quantum strategy can be represented as a coherent superposition over possible sabotage paths:
\begin{equation}
|\psi\rangle = \alpha |S_A\rangle + \beta |S_B\rangle,
\end{equation}
where $\alpha, \beta \in \mathbb{C}$ and the normalization condition holds:
\begin{equation}
|\alpha|^2 + |\beta|^2 = 1.
\end{equation}

In the general EWL framework, a quantum player's strategic action is modeled as a single-qubit unitary operation from the SU(2) group:
\begin{equation}
U(\theta, \phi) =
\begin{bmatrix}
\cos(\theta/2) & -e^{i\phi}\sin(\theta/2) \\
e^{-i\phi}\sin(\theta/2) & \cos(\theta/2)
\end{bmatrix},
\end{equation}
where $\theta \in [0, \pi]$ and $\phi \in [0, 2\pi)$ define the player's continuous strategy. In this study, however, we fix these local operations to identity (measurement in the computational basis) to isolate the contribution of the entangled resource state itself.

For multi-agent quantum sabotage games, entanglement among team members is realized through multipartite states. The simplest form of bipartite entanglement is the Bell state,
\begin{equation}
|\Phi^+\rangle = \frac{1}{\sqrt{2}}(|00\rangle + |11\rangle),
\end{equation}
which encodes perfect pairwise correlation between two agents. To generalize coordination across multiple players, we employ the W-state:
\begin{equation}
|W_N\rangle = \frac{1}{\sqrt{N}} \left( |10\ldots0\rangle + |010\ldots0\rangle + \cdots + |0\ldots01\rangle \right),
\end{equation}
which represents the equal superposition of all single-excitation configurations among $N$ qubits.

W-states preserve nontrivial multipartite correlations under certain local noise mechanisms and retain partial entanglement under qubit loss, making them useful for modeling distributed coordination in noisy settings. While Bell states capture pairwise correlations, W-states capture collective sharing of a single excitation across all agents, producing a characteristic joint-action distribution under measurement.

This formulation enables a unified description of quantum strategies by combining SU(2)-parameterized unitaries with multipartite entangled initial states. In this paper, we use this framework to compare induced joint-action statistics under identical information constraints.

\subsubsection{The Hybrid Adaptive Heuristic (HAH) Benchmark}

In addition to circuit-based strategies, we include a benchmark model designated as the Hybrid Adaptive Heuristic (HAH). The label is motivated by two ideas: hybrid quantum-classical (HQC) algorithm structure in which a classical controller guides a quantum co-processor \cite{HQC_Chen_2021}, and adaptive heuristic strategies from game theory \cite{Young_Adaptive_Heuristics_2006}. Crucially, HAH is not simulated by a physically realizable quantum circuit in this paper, and it should not be interpreted as a realizable strategy or as a quantitative point of comparison with circuit-based results. We use HAH only as an interpretive coordination ceiling to illustrate how strong coordination could compress outcome variance under idealized assumptions. All claims and conclusions about the QSG in this paper are based on the circuit-based (measurement-generated) strategies and their noise models.

\subsection{Multipartite Entanglement and the Sabotage Operator}

The QSG relies on multipartite entanglement across qubit registers and on noise processes that affect the circuit prior to measurement. This section formalizes the resource state and the channel notation used to describe noise.

\paragraph{Multipartite Entanglement.}
In this work, qubits are treated as two-level systems with computational basis
states $\ket{0}$ and $\ket{1}$.
The multipartite resource state is the symmetric W-state over $N$ qubits:
\begin{equation}
\ket{W_N} = \frac{1}{\sqrt{N}} \sum_{j=1}^{N}
\ket{0}^{\otimes (j-1)} \ket{1}_j \ket{0}^{\otimes (N-j)}.
\end{equation}

\paragraph{Sabotage Operator.}
We use $\mathcal{S}$ as a compact notation for the effective \emph{quantum channel}, i.e., a completely positive and trace-preserving (CPTP) map, acting on the circuit state prior to measurement. In particular, for a density matrix $\rho$,

\begin{equation}
\rho \mapsto \mathcal{S}(\rho),
\end{equation}
where $\mathcal{S}$ is assumed to be a completely positive, trace-preserving (CPTP) map capturing the chosen noise model (for example depolarizing, dephasing, amplitude damping, or bit-flip channels) \cite{nielsen2010quantum}. A unitary map is included as the special case.

\begin{equation}
\mathcal{S}(\rho) = U \rho U^\dagger.
\end{equation}

In this study, $\mathcal{S}$ is not selected adaptively by an adversary during gameplay; it denotes the fixed effective channel instantiated either by standard noise models or by the hardware-calibrated noise model used in simulation.

\subsection{Payoff Functions, Resource Evolution, and Nash Equilibrium}

The evolution of team resources in sabotage-only scenarios follows a dynamic process that incorporates both teams' action samples and, in the quantum case, their induced correlations. Let $R_Q(t)$ and $R_C(t)$ represent the quantum and classical team resources at time step $t$. We model the expected resource degradation due to sabotage by summing over joint probabilities of measured bitstrings:
\begin{equation}
E[R_Q(t+1)] = E[R_Q(t)] - \sum_{\mathbf{b} \in \{0,1\}^N} P_Q(\mathbf{b}) \sum_{k=1}^N S(b_k),
\end{equation}
\begin{equation}
E[R_C(t+1)] = E[R_C(t)] - \sum_{\mathbf{b} \in \{0,1\}^N} P_C(\mathbf{b}) \sum_{k=1}^N S(b_k).
\end{equation}

Here, $P(\mathbf{b})$ denotes the joint probability of measuring the bitstring $\mathbf{b} = (b_1, \dots, b_N)$, and $S(b_k)$ represents the score contribution associated with the $k$-th agent (determined by bit $b_k$). This formulation explicitly accounts for correlations in $\mathbf{b}$ rather than treating agents as independent marginals.

The expected utility of a quantum sabotage strategy $|\psi\rangle$ is
\begin{equation}
E[U_Q] = \sum_{i,j} P_{\text{measure}}(i,j) \cdot U(i,j),
\end{equation}
where $P_{\text{measure}}(i,j)$ is the joint probability of measuring sabotage actions $(i,j)$ from the quantum state and $U(i,j)$ is the utility function. In this work, we do not attempt to derive closed-form quantum Nash equilibria. Instead, we use the EWL notation to situate the model, while our primary focus is the induced joint-action statistics under fixed state preparation and fixed action mapping.

\section{Simulation Implementation}\label{sec:simulation_implementation}

This section describes the simulation implementation of the Quantum Sabotage Game (QSG) using IBM Qiskit \cite{cross2018ibm} and quantum simulation techniques. We compare classical and quantum strategies under four conditions: a non-circuit benchmark (HAH), ideal (noise-free) circuit simulation, circuit simulation with standard noise models, and circuit simulation using hardware-calibrated noise obtained via Qiskit Aer backends.

\subsection{Game Rules and Mechanics}

The Quantum Sabotage Game (QSG) simulates a competitive interaction between two opposing teams: a classical team (CT) and a quantum team (QT). The setting involves an army that exogenously defends one of two underground basements in each round, while players from both teams attempt to sabotage these strategic targets.

Each game round begins with a defense assignment. One of the two basements, denoted as \textit{A} and \textit{B}, is randomly marked as \textit{Strong} (defended), while the other is left as \textit{None} (undefended). The defense assignment is exogenous and passive; no player or team performs a defensive action, and defense only enters the game through the scoring of sabotage attempts. This assignment is hidden from the players. To ensure a fair comparison based on team size, we model two classical teams: a two-player team (2C) and a three-player team (3C). In both configurations, the agents each choose a sabotage path (\textit{A} or \textit{B}) without communication or coordination. Their actions are sampled randomly, reflecting decentralized decision-making typical of independent classical models.

In contrast, the quantum team makes decisions based on shared entangled quantum states. To explore how the structure of entanglement affects the induced joint-action distribution, we evaluate two quantum team configurations. The first is the Bell-State Team (BT), composed of two players (2Q) sharing a bipartite Bell state, compared against the 2C team. The second is the W-State Team (WT), consisting of three players (3Q) entangled via a multipartite W-state, compared against the 3C team. For the quantum teams, sabotage actions are determined by a single joint measurement of the shared entangled state. Each agent's action is set by the measurement outcome of its corresponding qubit using the fixed mapping $1\mapsto A$ and $0\mapsto B$. Team scores are computed by summing individual outcomes, with $+1$ assigned for a successful sabotage and $-1$ for a failed attempt.

This setup allows us to isolate how entanglement reshapes the joint-action distribution under identical information constraints, and to test how that correlation signature degrades under noise.

\subsection{Effectiveness Calculation and Ranking System}
To evaluate the success of sabotage and defense actions, we employ a standardized scoring system, summarized in Table \ref{tab:effectiveness_ranking}. It is worth noting that quantum teams leverage entanglement to alter the joint distribution of actions, which can be reflected in the distribution of round scores.

\begin{table}[!t]
\centering
\caption{Scoring rule and action-generation mechanisms in the Quantum Sabotage Game.}
\label{tab:effectiveness_ranking}
\footnotesize
\setlength{\tabcolsep}{3pt}
\renewcommand{\arraystretch}{1.1}
\begin{tabularx}{\columnwidth}{@{}l X c@{}}
\toprule
\textbf{Item} & \textbf{Definition} & \textbf{Value} \\
\midrule
Successful sabotage & Attack on the undefended basement (defense assigned exogenously) & $+1$ \\
Failed sabotage & Attack on the defended basement (hidden from players) & $-1$ \\
Classical player action & Independent choice of $A$ or $B$ without coordination & Individually scored \\
Bell-state team (2Q) & Bitwise-correlated measurement outcomes mapped to actions & Correlated actions \\
W-state team (3Q) & Multipartite single-excitation correlations mapped to actions & Correlated, non-identical actions \\
\bottomrule
\end{tabularx}
\end{table}

\subsection{HAH Benchmark vs. Classical Strategy}

\begin{figure}[!t]
  \centering
  \includegraphics[width=\columnwidth]{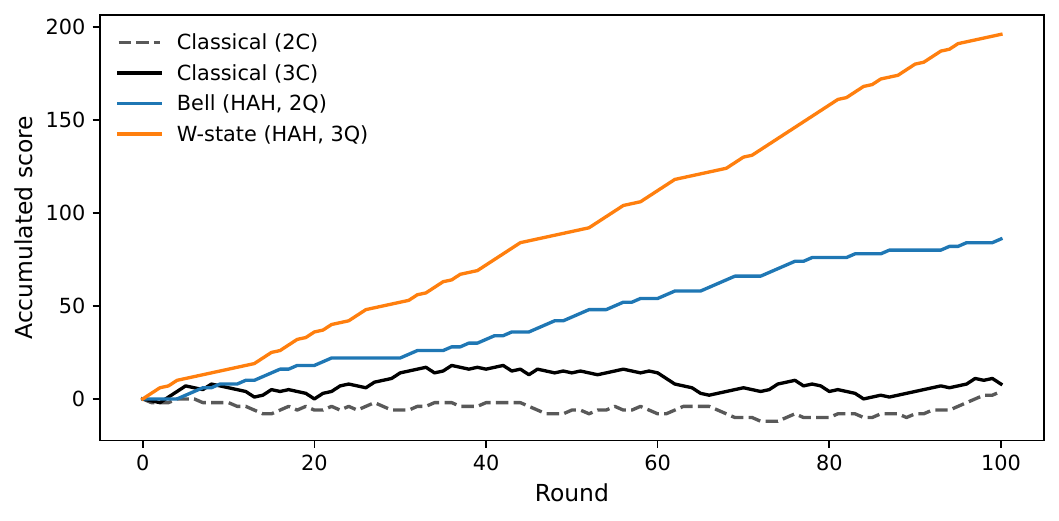}
 \caption{Accumulated sabotage scores over multiple rounds comparing classical teams with an idealized Hybrid Adaptive Heuristic (HAH) reference. The classical teams (2 players, 2C and 3 players, 3C) follow independent sabotage choices and fluctuate around zero. HAH is not a physically realizable strategy and is not used for quantitative comparison; it is included solely as an interpretive coordination ceiling to illustrate how strong coordination assumptions reshape finite-run trajectories without implying an expectation-value advantage.}

  \label{fig:classical_accumulated}
\end{figure}

\begin{figure}[!t]
  \centering
  \includegraphics[width=\columnwidth]{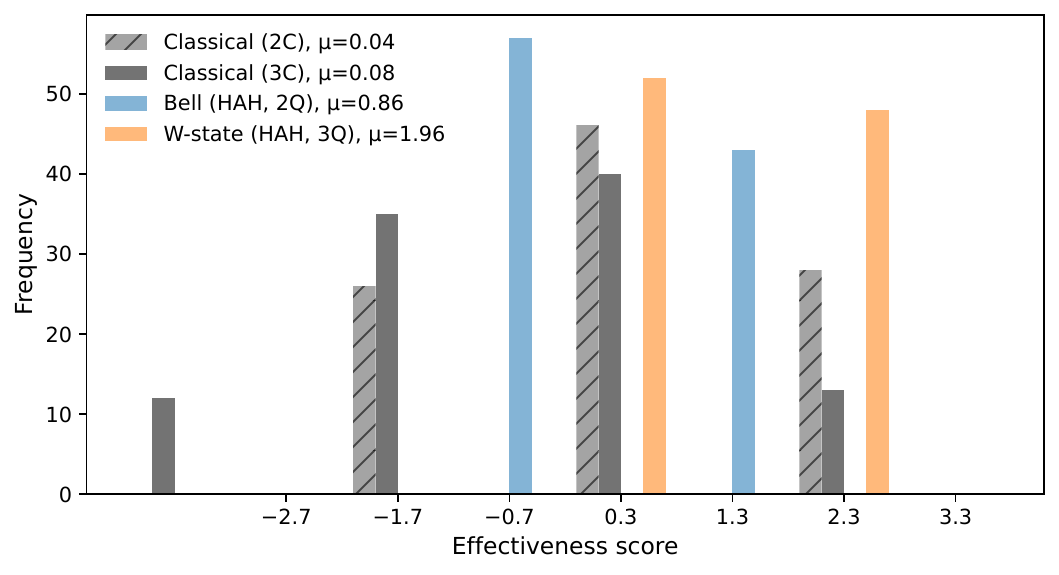}
    \caption{Distribution of sabotage effectiveness scores across classical teams and an idealized Hybrid Adaptive Heuristic (HAH) reference. HAH illustrates how externally imposed, non-physical coordination assumptions can suppress variance and reshape score distributions. Its reported sample mean is not drawn from the same stochastic process as the circuit-based strategies and therefore carries no implication for expectation values in the symmetric, hidden-defense game.}

  \label{fig:classical_effectiveness}
\end{figure}

The simulation was run for 100 rounds. We first include the Hybrid Adaptive Heuristic (HAH) benchmark, which represents an idealized coordination ceiling rather than a physically realizable circuit-based strategy (see Section 2.3.3). The HAH benchmark is not simulated using a direct quantum circuit, and its numerical values are not used as quantitative evidence for circuit-based conclusions. It is shown solely to illustrate how strong coordination assumptions can compress variance relative to independent sampling. The resulting effectiveness scores and accumulated scores over rounds for this benchmark are plotted in Figures \ref{fig:classical_accumulated} and \ref{fig:classical_effectiveness}. These curves are shown for qualitative illustration only and are not used for quantitative comparison with circuit-based strategies.
For clarity, any sample means shown in legends for this benchmark are finite-run descriptive statistics only, and they should not be interpreted as expectation-value comparisons with circuit-based strategies.

\subsection{Quantum Circuit Strategy Simulations}

We next simulate the quantum teams using Qiskit circuits. This "Pure Measurement" strategy generates actions directly from circuit measurement outcomes rather than from adaptive rules. We analyze three simulation tiers: ideal (noise-free), standard noise models, and hardware-calibrated noise.

A key modeling choice is how a single round samples a single joint action.
In our model, each round corresponds to one joint measurement of the shared
state, producing a single bitstring and therefore one action profile for the
team. Accordingly, in each round we execute the circuit with \texttt{shots=1}
(with \texttt{memory} enabled) and use the resulting bitstring directly to
assign each agent's action via the fixed mapping $1\mapsto A$ and $0\mapsto B$.

The 3-qubit W-state circuit was constructed using an efficient RY-CX cascade with fixed rotation angles $\theta_1 = 2\arccos(1/\sqrt{3})$ and $\theta_2 = 2\arccos(1/\sqrt{2})$, which deterministically prepares the symmetric state $|W_3\rangle = (|001\rangle + |010\rangle + |100\rangle)/\sqrt{3}$. This state is a specific instance of a single-excitation Dicke state, $D(3,1)$, and similar expansion preparation methods using such cascades are a subject of current research \cite{thapa2025expanding}. This compact realization reduces circuit depth and improves robustness.

\subsubsection{Ideal (Noise-Free) Conditions}

\begin{figure}[!t]
    \centering
    \includegraphics[width=\columnwidth]{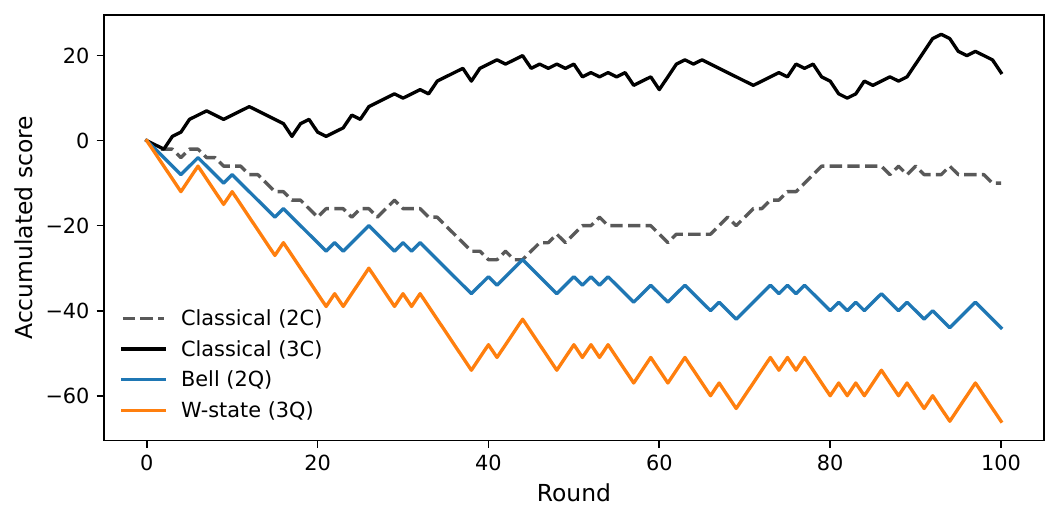}
    \caption{Accumulated sabotage scores over 100 rounds using pure Qiskit circuit strategies under ideal (noise-free) conditions. The 3Q W-State and 2Q Bell-State teams show distinct trajectories compared to the 3C and 2C classical teams, which fluctuate around zero.}
    \label{fig:realworld_accumulated}
\end{figure}

\begin{figure}[!t]
    \centering
    \includegraphics[width=\columnwidth]{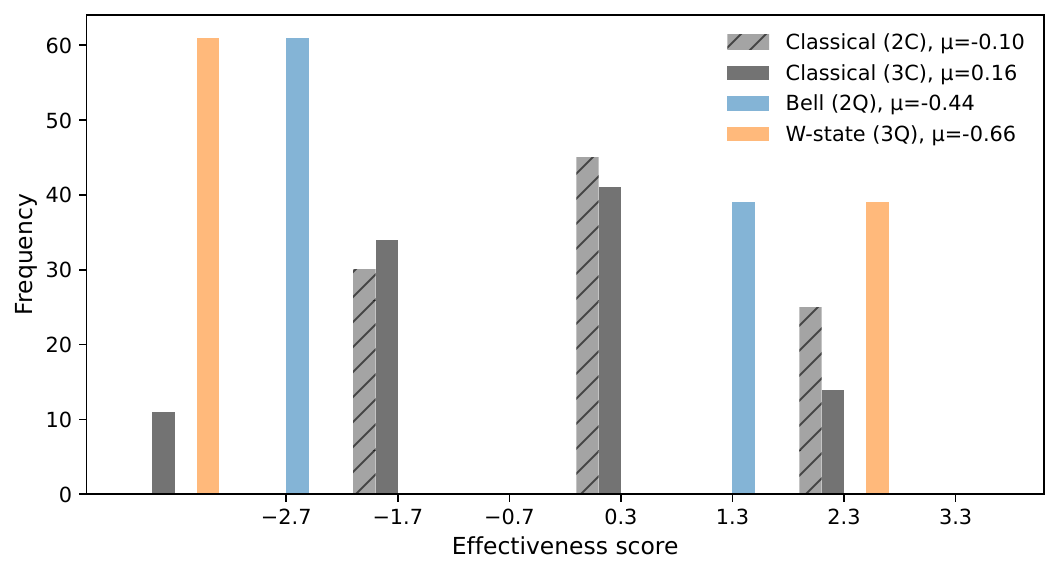}
    \caption{Effectiveness score distribution of pure Qiskit circuit strategies under ideal conditions. This histogram quantifies the distributional differences seen in Fig.~\ref{fig:realworld_accumulated}. The legend reports finite-run sample means for context.}
    \label{fig:realworld_effectiveness}
\end{figure}

First, we simulate the circuits under ideal noise-free conditions using the Qiskit Aer simulator. The accumulated sabotage scores and effectiveness distributions are shown in Figures~\ref{fig:realworld_accumulated} and~\ref{fig:realworld_effectiveness}. The results confirm a distinct coordination-induced difference in outcome statistics: the Bell and W-state teams produce joint-action distributions that differ from independent classical sampling under identical information constraints. Any finite-run sample means reported in legends are included only as descriptive statistics over a 100-round horizon and are not interpreted as asymptotic expected-utility shifts.

\subsubsection{Simulation with Standard Noise Models}

Having established the ideal-case coordination signature, we next evaluate the effects of standard quantum noise. We introduced depolarizing noise, amplitude-damping noise, and bit-flip errors into the quantum circuits. Each of these noise models captures different aspects of real-world imperfections in quantum hardware.
Depolarizing noise models the loss of coherence by replacing a quantum state with a maximally mixed state with probability $p$. The depolarizing channel for a single qubit is given by:
\begin{equation}
\mathcal{E}_{\text{depol}}(\rho) = (1 - p) \rho + \frac{p}{3} (X \rho X + Y \rho Y + Z \rho Z),
\end{equation}
where $X, Y,$ and $Z$ are the Pauli matrices, and $\rho$ is the density matrix of the system \cite{nielsen2010quantum}.

Amplitude-damping noise represents energy dissipation, such as photon loss in optical quantum systems. The Kraus operators for amplitude-damping noise are:
\begin{equation}
E_0 = \begin{bmatrix} 1 & 0 \\ 0 & \sqrt{1 - \gamma} \end{bmatrix}, \quad
E_1 = \begin{bmatrix} 0 & \sqrt{\gamma} \\ 0 & 0 \end{bmatrix}.
\end{equation}
Here, $\gamma$ represents the probability of energy loss.

Bit-flip noise introduces random flips of quantum bits, mimicking classical errors. The transformation is given by:
\begin{equation}
\mathcal{E}_{\text{bit-flip}}(\rho) = (1 - p) \rho + p X \rho X.
\end{equation}

\begin{figure}[!t]
    \centering
    \includegraphics[width=\columnwidth]{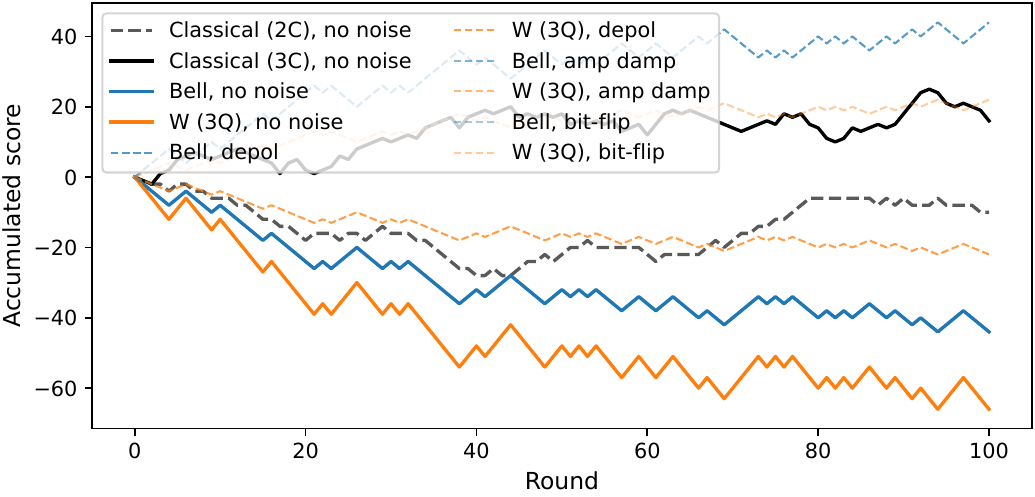}
    \caption{Accumulated score trajectories over 100 rounds for classical, Bell-state, and W-state teams under noise-free and standard noise conditions. The solid lines represent the ideal, noiseless scenario, while the dashed lines show behavior under depolarizing, amplitude-damping, and bit-flip noise. Under these noise models, the quantum teams' coordination signatures are reduced but can remain visible relative to classical baselines.}
    \label{fig:manual_noise_accumulated}
\end{figure}

\begin{figure}[!t]
    \centering
    \includegraphics[width=\columnwidth]{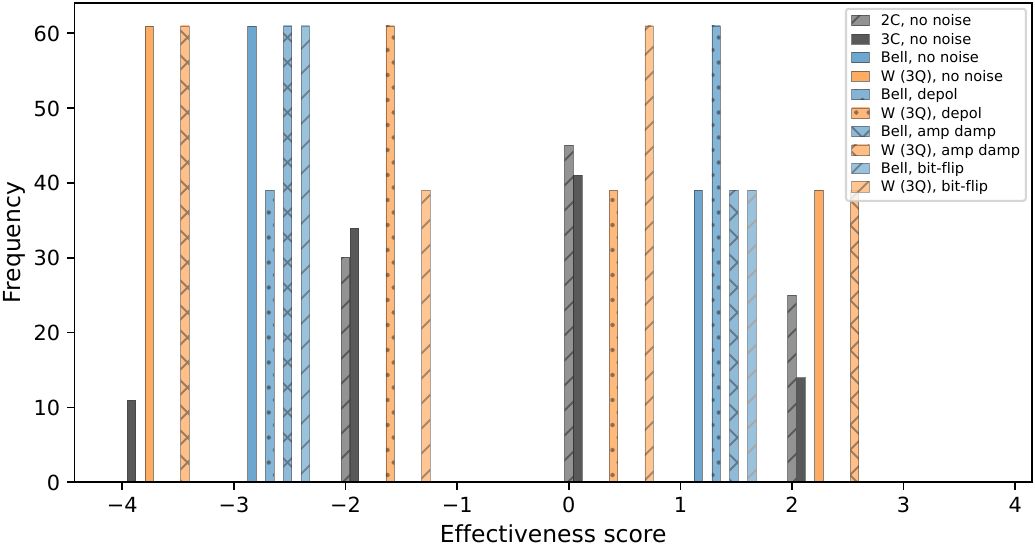}
   \caption{Effectiveness score distributions for classical, Bell-state, and W-state teams under noise-free conditions and under standard depolarizing, amplitude-damping, and bit-flip noise models. The noisy cases illustrate distributional compression and partial convergence toward classical-like sampling as decoherence increases. At the chosen error rate, these three canonical noise channels produce qualitatively similar aggregate distortion of the measurement-generated joint-action statistics, so the resulting histograms can appear closely overlapping. All reported means in this paper are finite-run descriptive statistics over a 100-round horizon, not asymptotic expected utilities.}

    \label{fig:manual_noise_effectiveness}
\end{figure}

By applying these noise models, we analyze how the coordination signature degrades. Figures \ref{fig:manual_noise_accumulated} and \ref{fig:manual_noise_effectiveness} illustrate the results under these noise models. The solid lines in Fig.~\ref{fig:manual_noise_accumulated} represent the ideal (noise-free) baseline, identical to Fig.~\ref{fig:realworld_accumulated}, while the dashed lines illustrate the effect of noise. All noise types progressively attenuate the entanglement-induced correlation signature, driving the joint-action statistics toward those of independent classical sampling. Nevertheless, under moderate noise levels and over finite horizons, the Bell-state and W-state teams can exhibit joint-action distributions that remain non-product, with observable differences in variance and tail behavior relative to classical baselines.

\subsubsection{Simulation with Real-World Hardware Noise}

To conclude our analysis, we simulated the sabotage game using a device-inspired, reproducible noise model provided by Qiskit Aer via the \textit{FakeKyiv} backend. Unlike manually constructed toy channels, this backend supplies a fixed set of gate and decoherence parameters intended to emulate representative NISQ-era behavior in a fully reproducible way. We use it as a consistent proxy for hardware-like noise rather than as a claim of matching any specific live device calibration.

\begin{figure}[!t]
    \centering
    \includegraphics[width=\columnwidth]{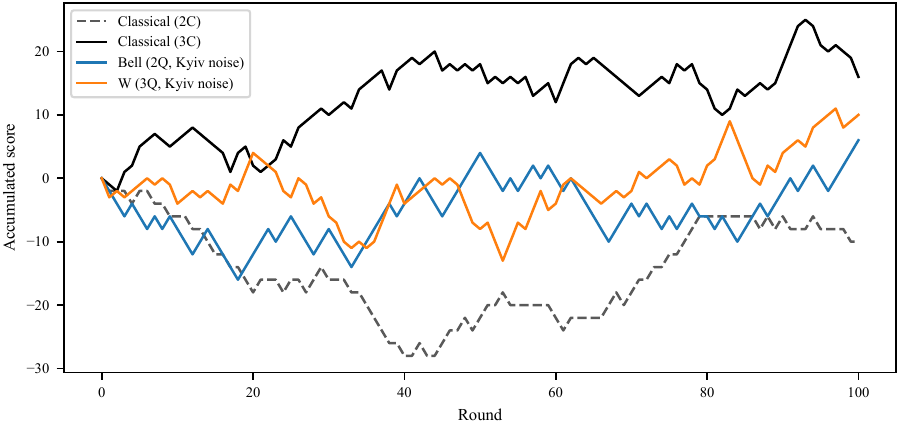}
    \caption{Accumulated sabotage scores over 100 rounds simulated with a hardware-calibrated
noise model (FakeKyiv). All strategies fluctuate around zero, consistent with the
symmetric payoff structure and hidden, exogenous defense assignment, which precludes
any asymptotic expected-score advantage. Differences in the accumulated trajectories
reflect finite-run fluctuations and, for the quantum teams, the presence of
measurement-induced correlations in joint actions rather than a systematic increase
in expected payoff. The Bell-state (2Q) and W-state (3Q) strategies therefore illustrate
distinct correlation-driven dynamics under realistic noise, not guaranteed dominance
over classical teams.
}
  \label{fig:ibm_noise_accumulated}
\end{figure}

\begin{figure}[!t]
    \centering
    \includegraphics[width=\columnwidth]{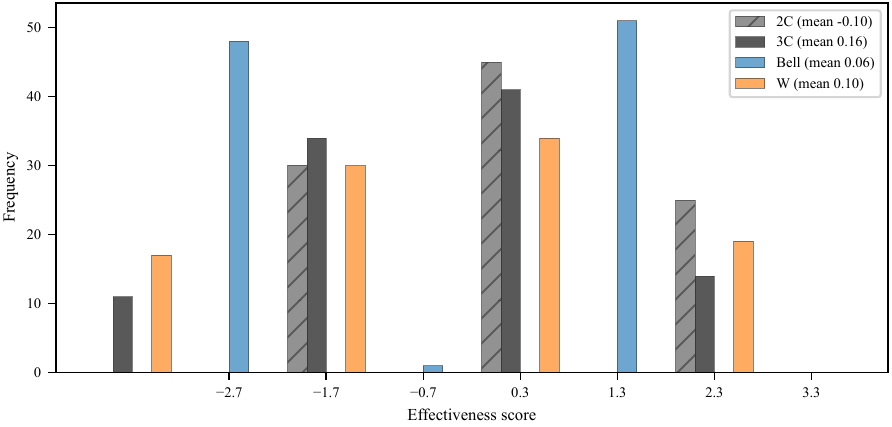}
    \caption{Distribution of per-round effectiveness scores under hardware-calibrated noise (\protect\textit{FakeKyiv}) for classical and quantum teams. While all strategies have zero asymptotic expected score by symmetry, the quantum teams exhibit distinct distributional signatures arising from entanglement-induced correlations. In the ideal case, the Bell-state (2Q) strategy is restricted to even-valued outcomes, while under noise odd-valued outcomes can appear as correlations degrade, whereas the W-state (3Q) strategy produces characteristic single-excitation outcomes ($\pm 1$ and $\pm 3$), reflecting multipartite correlation structure. Under realistic noise, these nonclassical distributional features are partially degraded but remain observable, demonstrating the persistence of entanglement-induced coordination signatures beyond independent classical sampling.}
    \label{fig:ibm_noise_effectiveness}
\end{figure}

Figures \ref{fig:ibm_noise_accumulated} and \ref{fig:ibm_noise_effectiveness} illustrate the accumulated score and effectiveness distribution under this hardware-calibrated noise. These results demonstrate a key finding: even under realistic noise, both the Bell-state (2Q) and W-state (3Q) teams can maintain outcome distributions that differ from their size-equivalent classical counterparts (2C and 3C) in a finite-run evaluation. Any sample means shown in legends are descriptive statistics over the 100-round horizon and are interpreted here as reflecting distribution reshaping and correlation persistence rather than an expectation-value guarantee.

\section{Results}\label{sec:results}

This section presents the outcomes of our simulations, comparing classical and quantum sabotage strategies under various conditions. We analyze accumulated scores, effectiveness distributions, and the impact of quantum noise to characterize coordination signatures induced by entanglement.

All reported mean values are finite-sample statistics over 100 rounds and should not be interpreted as asymptotic expected utilities; deviations from zero in short horizons can occur due to sampling fluctuations, finite-run drift, and distribution reshaping, not because the symmetric game guarantees a positive expected payoff.

We first include the Hybrid Adaptive Heuristic (HAH) benchmark (Figures~\ref{fig:classical_accumulated} and~\ref{fig:classical_effectiveness}) as an idealized coordination ceiling. As emphasized earlier, HAH is not physically realizable as a circuit-based strategy in this paper, and its numerical values are not used as quantitative evidence for circuit-based conclusions. Its role is interpretive: it illustrates how strong coordination assumptions reshape score distributions, reducing variance relative to independent sampling, without implying an achievable expected-utility gain.

We then evaluate the circuit-based ``Pure Measurement'' strategy using Qiskit circuits in an ideal, noise-free environment. Figure~\ref{fig:realworld_accumulated} shows that Bell- and W-state teams generate accumulated score trajectories that differ from independent classical sampling over a finite 100-round horizon, while Figure~\ref{fig:realworld_effectiveness} shows the corresponding differences in round-score distributions. Any sample means reported in the legends are finite-run descriptive statistics included for context only. At this stage, the emphasis is on how entanglement reshapes the joint-action distribution relative to independent classical sampling, rather than on expectation-value shifts, which are ruled out by symmetry under hidden, uniform defense.

Under a uniform random defense assignment (Strong on $A$ or $B$ with equal probability) that is hidden from the players, any strategy that does not condition on the defense state has zero asymptotic expected score. Nevertheless, over finite simulation runs (100 rounds), accumulated scores can exhibit nonzero trajectories and nonzero sample means. This occurs because entanglement-based coordination reshapes the joint-action distribution, which can suppress or amplify particular joint outcomes and thereby alter finite-run variance, tail behavior, and apparent drift in a short horizon. The reported sample means are therefore finite-run descriptive statistics reflecting distributional reshaping and sampling fluctuations, not a guaranteed expectation-value advantage. When averaged over many independent 100-round runs, the sample mean converges to zero as required by symmetry; any nonzero mean in a single displayed run is a finite-horizon fluctuation rather than an expectation-value advantage.

We next apply three standard noise models: depolarizing noise,
amplitude-damping noise, and bit-flip errors.
Figure~\ref{fig:manual_noise_accumulated} shows that noise erodes the coordination
signature, with dashed (noisy) trajectories converging toward the classical
baseline.
Figure~\ref{fig:manual_noise_effectiveness} shows how noise redistributes weight
across outcomes and compresses the score distribution toward classical-like
sampling.

Finally, we simulate the QSG using a calibrated noise model provided by Qiskit Aer
via the \textit{FakeKyiv} backend.
Figure~\ref{fig:ibm_noise_accumulated} shows that quantum strategies can exhibit
finite-run deviations from independent classical sampling over the displayed
horizon, while Figure~\ref{fig:ibm_noise_effectiveness} shows that the
corresponding round-score distributions remain separated relative to classical
baselines.
In this setting, the qualitative conclusion is that a correlation signature in
joint-action sampling can persist under representative NISQ-era noise at small
team sizes.

To quantify this correlation structure independently of mean score, we evaluate
correlation-sensitive observables directly from the measurement outcomes.
For the Bell-state strategy, we define the two-body correlator
\begin{equation}
\langle Z_1 Z_2 \rangle = \sum_{b_1,b_2\in\{0,1\}} (-1)^{b_1+b_2}\,P(b_1 b_2),
\end{equation}
where $P(b_1 b_2)$ is the observed probability of measuring bitstring $b_1 b_2$.
This quantity equals $+1$ for perfectly correlated outcomes ($00$ or $11$) and
vanishes for independent unbiased sampling.

For the W-state strategy, we quantify multipartite coordination by the
single-excitation rate
\begin{equation}
p_{\mathrm{1exc}} = \sum_{\mathbf{b}:\,|\mathbf{b}|=1} P(\mathbf{b}),
\end{equation}
where $|\mathbf{b}|$ denotes the Hamming weight of the measured bitstring.
An ideal $|W_3\rangle$ state yields $p_{\mathrm{1exc}}=1$, whereas independent
unbiased sampling yields $p_{\mathrm{1exc}}=3/8$ for three bits.

Table~\ref{tab:effectiveness_comparison} summarizes finite-run sample means and
accumulated totals for the 100-round trajectories shown in the corresponding
figures.
These values are included only as compact descriptors of the displayed
finite-horizon runs.
To avoid misinterpretation, we emphasize that when averaged over many independent
100-round runs (or equivalently over independently resampled defense sequences),
the sample mean approaches zero for all strategies, as required by symmetry under
hidden, uniform defense.

\emph{Table~\ref{tab:effectiveness_comparison} reports descriptive statistics for
the specific 100-round trajectories shown in
Figs.~\ref{fig:realworld_accumulated}--\ref{fig:ibm_noise_effectiveness}
(one fixed seed) and is included only as a compact summary of those displayed
runs, not as an expected-utility comparison.}

\begin{table}[!t]
	\centering
	\caption{Single-run descriptive statistics for the displayed 100-round trajectories (not expected utilities).}
	\label{tab:effectiveness_comparison}
	\footnotesize
	\setlength{\tabcolsep}{3pt}
	\renewcommand{\arraystretch}{1.15}
	
	\begin{tabularx}{\columnwidth}{l l >{\centering\arraybackslash}X >{\centering\arraybackslash}X}
		\toprule
		\textbf{Strategy} & \textbf{Noise} &
		\textbf{Sample mean (one run)} &
		\textbf{Sample total (one run)} \\
		\midrule
		Classical (2C)     & Baseline & 0.04 & 4  \\
		Classical (3C)     & Baseline & 0.08 & 8  \\
		Quantum (Bell, 2Q) & Ideal    & 0.08 & 8  \\
		Quantum (W, 3Q)    & Ideal    & 0.30 & 30 \\
		Quantum (Bell, 2Q) & FakeKyiv & 0.24 & 24 \\
		Quantum (W, 3Q)    & FakeKyiv & 0.28 & 28 \\
		\bottomrule
	\end{tabularx}
	
\end{table}

From these results, several consistent trends emerge. Circuit-based quantum strategies generate joint-action distributions that differ from independent classical sampling in ideal conditions, with W-state entanglement producing a characteristic multipartite correlation structure. As noise increases, these signatures weaken, yet under representative hardware-calibrated noise, distributional differences can remain observable for small-scale entanglement (2--3 qubits) over finite horizons. The main point is therefore about correlation structure and distribution reshaping, not about guaranteed expectation-value improvement in a symmetric, hidden-defense setting.

\section{Discussion}\label{sec:discussion}

Our study of the Quantum Sabotage Game provides a model for analyzing adversarial interactions in the quantum domain, extending sabotage-only team games to include entanglement-generated joint-action sampling. The central finding is that multipartite entanglement, particularly W-states, can induce a correlation structure in decentralized actions that remains distinct from independent classical sampling under identical information constraints, and that this signature can persist under realistic noise.

The introduction of an explicit channel notation $\mathcal{S}$ within the EWL-style quantization vocabulary is intended to formalize how noise reshapes outcome statistics in an adversarial environment. In contrast to settings where quantum resources are framed as resolving dilemmas or increasing expected payoff, the present model is symmetric and defense is hidden and exogenous, so expectation-value claims are not the appropriate metric. Instead, the relevant observable is the joint-action distribution and its derived features, including variance, tail behavior, and correlation structure.

Our analysis reveals that the persistence of entanglement-induced coordination signatures is critically dependent on coherence. As noise increases, the joint-action distribution converges toward a classical-like product distribution, reducing observable distinctions. This behavior is consistent with prior studies showing that quantum game signatures are sensitive to physical noise mechanisms that modify underlying quantum correlations \cite{Ozaydin2016}. Here, the same principle appears in a sabotage-only, hidden-defense setting: as decoherence grows, correlation structure degrades and distributional distinctions compress.

These results suggest that if quantum-enhanced coordination is to be used in practical adversarial decision-making settings, noise-resilient state preparation and mitigation will be essential. At the same time, the simulations indicate that small-scale entanglement can retain observable correlation signatures under representative NISQ-era noise, motivating further studies of scalable state families and error mitigation in multi-agent quantum games.

\section{Conclusion}

In this study, we investigated the role of multipartite entanglement in adversarial settings through the formulation and simulation of the Quantum Sabotage Game. We showed that entangled resource states, particularly 3-qubit W-states, induce coordination signatures in decentralized actions that are not reproduced by independent classical sampling under identical information constraints.

We conducted simulations comparing classical, Bell-state, and W-state strategies under ideal conditions, standard noise models, and hardware-inspired noise instantiated via a reproducible Qiskit Aer backend. By structuring the analysis around size-equivalent teams (2C vs. 2Q and 3C vs. 3Q), we isolated how entanglement reshapes the joint-action distribution. In ideal environments, the quantum teams exhibited structured correlations distinct from classical teams. This distinction narrows as noise increases, yet circuit-based quantum strategies can remain distributionally distinct from classical baselines under representative NISQ-era noise for small team sizes over finite horizons.

Overall, this work provides a framework for studying how quantum correlations influence adversarial multi-agent decision-making in the presence of noise. The emphasis is on distributional and correlation signatures rather than on expectation-value advantage in symmetric hidden-information settings.


\appendices

\section{Simulation Environment and Reproducibility}

All simulations were conducted using Python scripts for reproducibility.
The key software packages were \texttt{Qiskit}, \texttt{Qiskit Aer},
\texttt{NumPy}, and \texttt{Matplotlib}.

Reproducibility was enforced by fixing random seeds at three levels.
A NumPy seed (\texttt{20251021}) controls all classical randomization, including
the exogenous defense assignment and the independent classical team actions.
A Qiskit transpiler seed (\texttt{20251021}) fixes the compilation and routing
choices so that circuit optimization is identical across runs.
For circuit execution, reproducibility is handled differently for ideal and noisy
simulations. In the ideal (noise-free) case, an Aer simulator seed is fixed to make the pseudo-random sampling reproducible across runs. In simulations that include noise channels (standard toy noise models or the device-inspired
\textit{FakeKyiv} noise model), the simulator seed is decorrelated across rounds
(e.g., $\texttt{seed\_simulator}=\texttt{seed}_0+t$ for round $t$) to avoid
artificially locking the one-shot ($\texttt{shots}=1$) sampling sequence. In all
cases, the transpiler seed is fixed so that circuit compilation and routing are
identical across runs.

In each round, the corresponding quantum circuit is executed with
\texttt{shots=1} (with \texttt{memory} enabled), producing a single measured
bitstring that directly defines the joint action profile for that round via the
fixed mapping $1\mapsto A$ and $0\mapsto B$.

\section{Team Definitions}

All simulations compared four distinct teams over 100 rounds. The team sizes were fixed to ensure fair, size-equivalent comparisons: a two-player Classical Team ($\mathbf{N=2C}$), a three-player Classical Team ($\mathbf{N=3C}$), a two-qubit Bell-State Team ($\mathbf{N=2Q}$), and a three-qubit W-State Team ($\mathbf{N=3Q}$).

\section{Quantum State Preparation Circuits}

\begin{figure}[h]
    \centering
    \includegraphics[width=\columnwidth]{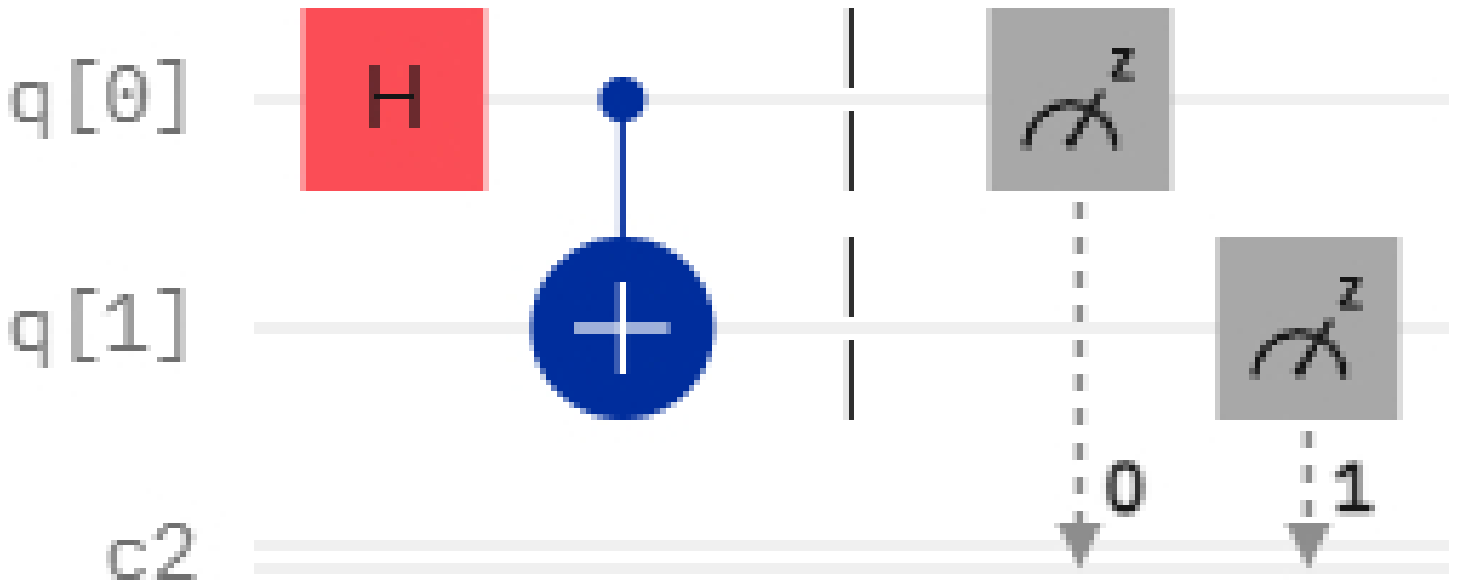}
    \caption{Circuit used to prepare the 2-qubit Bell state.}
    \label{fig:S1}
\end{figure}

The 2Q and 3Q quantum teams used circuits to prepare their entangled states, as shown in Figures \ref{fig:S1} and \ref{fig:S2}.

\subsubsection{ Bell-State (2Q) Circuit}
The 2Q team used the standard circuit (Hadamard on qubit 0, CNOT from 0 to 1) to prepare the $|\Phi^+\rangle$ state.

\subsubsection{W-State (3Q) Circuit}
\begin{figure}[h]
    \centering
    \includegraphics[width=\columnwidth]{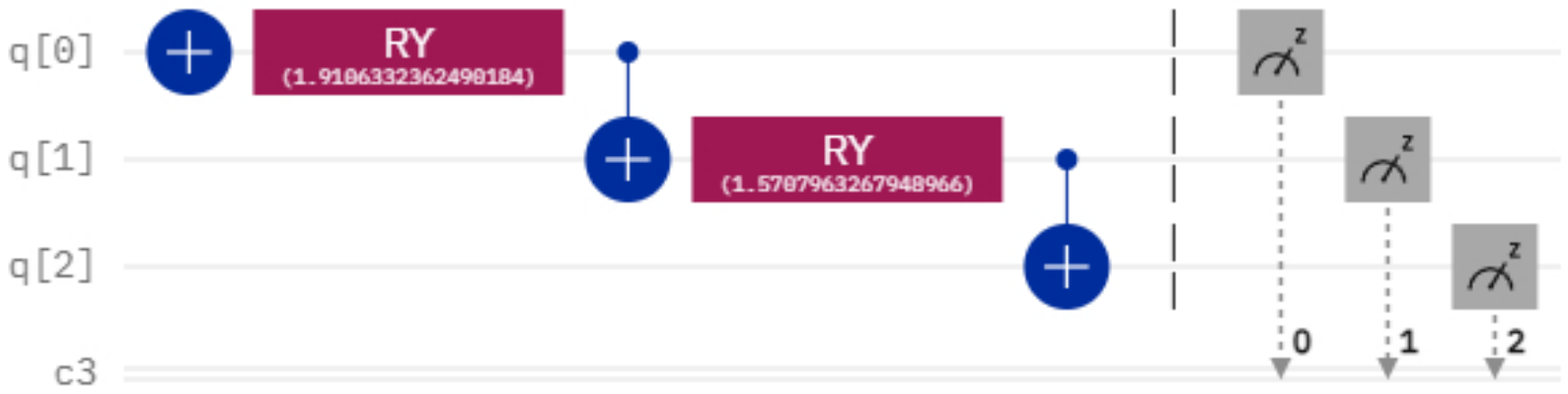}
    \caption{Circuit used for deterministic preparation of the 3-qubit W-state, showing the specific rotation angles.}
    \label{fig:S2}
\end{figure}

The 3Q team used an efficient, deterministic circuit to prepare the $|W_3\rangle$ state, as shown in Fig. \ref{fig:S2}. This circuit starts with an \texttt{X} gate on qubit 0 (to create $|100\rangle$) and then applies a cascade of \texttt{RY} rotations and \texttt{CNOT} gates. The rotation angles ($\theta_i$) are calculated analytically using the formula:
\[ \theta_i = 2 \arccos\left(\sqrt{\frac{N-i-1}{N-i}}\right) \]
For $N=3$ qubits, this yields two angles: for $i=0$, the angle is $\theta_0 = 2 \arccos(\sqrt{2/3}) \approx 1.2310$ rad, and for $i=1$, the angle is $\theta_1 = 2 \arccos(\sqrt{1/2}) = \pi/2 \approx 1.5708$ rad.

\section{Game Logic and Measurement Sampling}

A critical aspect of the simulation logic is the method for determining the
quantum team's action in each round. For a given round, the appropriate quantum
circuit (Bell-state or W-state) is executed on an Aer-based simulator backend with
\texttt{shots=1} (with the \texttt{memory} setting enabled), producing a single
measured bitstring. This single joint measurement outcome directly defines the
individual actions of all team members for that round via the fixed mapping
$\texttt{'1'} \rightarrow \texttt{A}$ and $\texttt{'0'} \rightarrow \texttt{B}$.

The total score for the round is then calculated by summing the individual scores (+1 for success, $-1$ for failure) for all players on that team. This procedure is repeated independently for 100 rounds, with each round corresponding to one realized joint measurement and one realized team action profile.

\section{Simulation Scenario Definitions}

The manuscript presents data from four distinct simulations.

The first simulation is the HAH Benchmark. This scenario does not simulate quantum circuits. Instead, it compares the two classical teams (2C, 3C) against rule-based proxy functions that model perfect, entanglement-assisted classical adaptation (see Sec. 2.3.3 of the main text).

The second scenario is the Ideal (Noise-Free) Circuit Simulation. This simulation uses the Qiskit Aer simulator with no noise model. It simulates the
``Pure Measurement'' strategy, establishing the baseline
entanglement-induced coordination signature generated by the circuits alone.

The third scenario is the Standard Noise Model (SNM) Simulation. This uses the same logic as the ideal simulation but injects three different standard noise models from Qiskit Aer. These include depolarizing noise, amplitude damping noise, and bit-flip noise, each applied with a 5\% error rate.

The fourth and final scenario is the Hardware Noise Simulation. This loads a realistic, calibrated noise model from a reproducible Qiskit Aer backend (\textit{FakeKyiv}) and passes the resulting noise model object to the simulator, providing a robust, reproducible simulation of how the quantum strategies would perform under representative NISQ-era noise.

\EOD

\begin{thebibliography}{99}


\bibitem{von1947theory}
J.~Von Neumann and O.~Morgenstern,
\textit{Theory of games and economic behavior, 2nd rev},
Princeton University Press, 1947.

\bibitem{nash1950equilibrium}
J.~F. Nash Jr.,
``Equilibrium points in n-person games,''
\textit{Proceedings of the National Academy of Sciences},
vol.~36, no.~1, pp.~48--49, 1950.

\bibitem{osborne1994course}
M.~J. Osborne and A.~Rubinstein,
\textit{A course in game theory},
MIT Press, Boston, USA, 1994.

\bibitem{patil2018applications}
A.~P. Patil, S.~Bharath, and N.~M. Annigeri,
``Applications of game theory for cyber security system: A survey,''
\textit{International Journal of Applied Engineering Research},
vol.~13, no.~17, pp.~12987--12990, 2018.

\bibitem{leimar2023game}
O.~Leimar and J.~M. McNamara,
``Game theory in biology: 50 years and onwards,''
\textit{Philosophical Transactions of the Royal Society B},
vol.~378, no.~1876, p.~20210509, 2023.

\bibitem{ho2022game}
E.~Ho, A.~Rajagopalan, A.~Skvortsov, S.~Arulampalam, and M.~Piraveenan,
``Game theory in defence applications: A review,''
\textit{Sensors}, vol.~22, no.~3, p.~1032, 2022.

\bibitem{eisert1999quantum}
J.~Eisert, M.~Wilkens, and M.~Lewenstein,
``Quantum games and quantum strategies,''
\textit{Physical Review Letters}, vol.~83, no.~15, p.~3077, 1999.

\bibitem{meyer1999quantum}
D.~A. Meyer,
``Quantum strategies,''
\textit{Physical Review Letters}, vol.~82, no.~5, p.~1052, 1999.

\bibitem{piotrowski2003invitation}
E.~W. Piotrowski and J.~S{\l}adkowski,
``An invitation to quantum game theory,''
\textit{International Journal of Theoretical Physics}, vol.~42, pp.~1089--1099, 2003.

\bibitem{Flitney2002}
A.~P. Flitney and D.~Abbott,
``An introduction to quantum game theory,''
\textit{Fluctuation and Noise Letters}, vol.~2, no.~4, pp.~R175--R187, 2002.

\bibitem{Brunner2013}
N.~Brunner and N.~Linden,
``Connection between Bell nonlocality and Bayesian game theory,''
\textit{Nat. Commun.}, vol.~4, p.~2057, 2013.

\bibitem{Du2020}
J.~Du, X.~Xu, M.~Shi, J.~Wu, X.~Zhou, and R.~Han,
``Experimental realization of quantum games on a quantum computer,''
\textit{Phys. Rev. A}, vol.~81, no.~6, p.~064102, 2020.

\bibitem{Iqbal2009}
A.~Iqbal and D.~Abbott,
``Quantum Colonel Blotto game,''
\textit{Phys. Rev. A}, vol.~80, no.~3, p.~032328, 2009.

\bibitem{du2002entanglement}
J.~Du, H.~Li, X.~Xu, X~Zhou, and R~Han,
``Entanglement enhanced multiplayer quantum games,''
\textit{Physics Letters A}, vol.~302, no.~5--6, pp.~229--233, 2002.

\bibitem{li2022prisoner}
D.~Li, X.~Sun, Y.~He, and D.~Han,
``On prisoner’s dilemma game with psychological bias and memory learning,''
\textit{Applied Mathematics and Computation}, vol.~433, p.~127390, 2022.

\bibitem{altintas2022prisoners}
A.~A. Altintas, F.~Ozaydin, C.~Bayindir, and V.~Bayrakci,
``Prisoners’ dilemma in a spatially separated system based on spin--photon interactions,''
\textit{Photonics}, vol.~9, no.~9, p.~617, 2022.

\bibitem{lopez2012battle}
J.~L{\'o}pez \textit{et al.},
``Battle of the sexes game analysis using Yang-Baxter operator as quantum gate,''
in \textit{Quantum Information and Computation X}, vol.~8400, pp.~152--161, 2012. SPIE.

\bibitem{maioli2019quantization}
A.~C. Maioli, M.~H.~M. Passos, W.~F. Balthazar, C.~E.~R. Souza, J.~A.~O. Huguenin, and A.~G.~M. Schmidt,
``Quantization and experimental realization of the Colonel Blotto game,''
\textit{Quantum Information Processing}, vol.~18, pp.~1--20, 2019.

\bibitem{jayanti4712364multiplayer}
S.~Jayanti,
``The Multiplayer Colonel Blotto Game on the Interval and Other Measure Spaces,''
\textit{SSRN} 4712364.

\bibitem{xu2022experimental}
J.-M. Xu, Y.-Z. Zhen, Y.-X. Yang, Z.-M. Cheng, Z.-C. Ren, K.~Chen, X.-L. Wang, and H.-T. Wang,
``Experimental demonstration of quantum pseudotelepathy,''
\textit{Physical Review Letters}, vol.~129, no.~5, p.~050402, 2022.

\bibitem{makram2024time}
A.~T.~M. Makram-Allah, M.~Y. Abd-Rabbou, and N.~Metwally,
``Time dependence of Eisert--Wilkens--Lewenstein quantum game,''
\textit{Quantum Information Processing}, vol.~23, no.~12, p.~393, 2024.

\bibitem{sanz2025mapping}
L.~Sanz-Mart{\'\i}n, G.~Rivas, N.~Clavijo-Buritic{\'a}, M.~Herrera, and J.~Parra-Dom{\'\i}nguez,
``Mapping the frontier: a review of quantum and evolutionary game theory for complex decision-making,''
\textit{Quantum Information Processing}, vol.~24, no.~9, pp.~1--34, 2025.

\bibitem{shen2025quantum}
J.~Shen, L.~Shi, and K.~Zhu,
``A quantum game model for dual channels under channel cooperation and service: J. Shen et al.,''
\textit{Quantum Information Processing}, vol.~24, no.~10, p.~307, 2025.

\bibitem{tiago2025classical}
G.~Tiago, J.~Naskar, A.~C. Maioli, W.~F. Balthazar, A.~G.~M. Schmidt, and J.~A.~O. Huguenin,
``Classical and quantum multiplayer Colonel Blotto game in all-optical setup: G. Tiago et al.,''
\textit{Quantum Information Processing}, vol.~24, no.~5, p.~152, 2025.

\bibitem{schmid2010experimental}
C.~Schmid, A.~P. Flitney, W.~Wieczorek, N.~Kiesel, H.~Weinfurter, and L.~C.~L. Hollenberg,
``Experimental implementation of a four-player quantum game,''
\textit{New J. Phys.}, vol.~12, no.~6, p.~063031, 2010.

\bibitem{Zhao2021}
L.~Zhao, X.~Liu, and J.~Chen,
``Quantum sabotage games in secure communication networks,''
\textit{Quantum Rep.}, vol.~3, no.~2, pp.~219--234, 2021.

\bibitem{Gross2020}
J.~Gross and E.~Wagner,
``Sabotage games in strategic resource allocation,''
\textit{Game Theory Review}, vol.~6, no.~3, pp.~189--204, 2020.

\bibitem{harsanyi1973games}
J.~C. Harsanyi,
``Games with randomly disturbed payoffs: A new rationale for mixed-strategy equilibrium points,''
\textit{International Journal of Game Theory}, vol.~2, no.~1, pp.~1--23, 1973.

\bibitem{levine2017deep}
Y.~Levine, D.~Yakira, N.~Cohen, and A.~Shashua,
``Deep learning and quantum entanglement: Fundamental connections with implications to network design,''
\textit{arXiv preprint} arXiv:1704.01552, 2017.

\bibitem{kolokoltsov2022dynamic}
V.~N. Kolokoltsov,
``Dynamic quantum games,''
\textit{Dynamic Games and Applications}, vol.~12, no.~2, pp.~552--573, 2022.

\bibitem{Bugu2020}
S.~Bugu, F.~Ozaydin, and T.~Kodera,
``Surpassing the classical limit in magic square game with distant quantum dots coupled to optical cavities,''
\textit{Scientific Reports}, vol.~10, p.~22202, 2020.

\bibitem{Ozdemir2011}
S.~K. Ozdemir, E.~Matsunaga, T.~Tashima, T.~Yamamoto, M.~Koashi, and N.~Imoto,
``An Optical Fusion Gate for {W}-States,''
\textit{New J. Phys}, vol.~13, p.~103003, 2011.

\bibitem{Bugu2013A}
S.~Bugu, C.~Yesilyurt, and F.~Ozaydin,
``Enhancing the {W}-State Quantum-Network-Fusion Process with a Single {F}redkin Gate,''
\textit{Phys. Rev. A}, vol.~87, p.~032331, 2013.

\bibitem{Ozaydin2014A}
F.~Ozaydin, S.~Bugu, C.~Yesilyurt, A.~A. Altintas, M.~Tame, and S.~K. Ozdemir,
``Fusing Multiple {W} States Simultaneously with a {F}redkin Gate,''
\textit{Phys. Rev. A}, vol.~89, p.~042311, 2014.

\bibitem{thapa2025expanding}
Thapa, B., Moran, O., Vu, D.-K., \& Ozaydin, F. (2025). Expanding a 4-qubit Dicke State to a 5-qubit Dicke State with Limited Qubit Access. \textit{arXiv preprint arXiv:2508.07977}.

\bibitem{nielsen2010quantum}
M.~A. Nielsen and I.~L. Chuang,
\textit{Quantum computation and quantum information},
Cambridge University Press, Cambridge, UK, 2010.

\bibitem{zhu2024robustness}
L.-H. Zhu, Z.~Zhu, G.-L. Lv, C.-Q. Ye, and X.-Y. Chen,
``Robustness of Entanglement for Dicke-W and Greenberger-Horne-Zeilinger Mixed States,''
\textit{Entropy}, vol.~26, no.~9, p.~804, 2024.

\bibitem{preskill2018quantum}
J.~Preskill,
``Quantum computing in the NISQ era and beyond,''
\textit{Quantum}, vol.~2, p.~79, 2018.

\bibitem{tambe2011security}
M.~Tambe,
\textit{Security and game theory: algorithms, deployed systems, lessons learned},
Cambridge University Press, Cambridge, UK, 2011.


\bibitem{Ozaydin2016}
F.~Ozaydin,
``Effect of Dzyaloshinskii–Moriya interaction on the winning probability of the magic square game,''
\textit{Quantum Inf. Process.}, vol.~15, no.~7, pp.~2531--2541, 2016.

\bibitem{HQC_Chen_2021}
R.-Y.-L. Chen, B.-C. Zhao, Z.-X. Song, X.-Q. Zhao, K.~Wang, and X.~Wang,
``Hybrid quantum-classical algorithms: Foundation, design and applications,''
\textit{Acta Physica Sinica}, vol.~70, no.~21, p.~210302, 2021.

\bibitem{Young_Adaptive_Heuristics_2006}
H.~P. Young,
\textit{Adaptive Heuristics},
Technical Report, Johns Hopkins University, Department of Economics, 2006.
Available at: \url{http://www.econ2.jhu.edu/People/Young/Adaptiveheuristics.pdf}.

\bibitem{callison2022hybrid}
A.~Callison and N.~Chancellor,
``Hybrid quantum-classical algorithms in the noisy intermediate-scale quantum era and beyond,''
\textit{Physical Review A}, vol.~106, no.~1, p.~010101, 2022.

\bibitem{Guerreschi2017PracticalOF}
G.~G. Guerreschi and M.~Smelyanskiy,
``Practical optimization for hybrid quantum-classical algorithms,''
\textit{arXiv: Quantum Physics}, 2017.

\bibitem{cross2018ibm}
A.~Cross,
``The IBM Q experience and QISKit open-source quantum computing software,''
\textit{APS March Meeting Abstracts}, vol.~2018, pp.~L58--003, 2018.

\end{thebibliography}
\end{document}